# Design of a High Speed XAUI Based on Dynamic Reconfigurable Transceiver IP Core

*,[1]Haipeng Zhang, [1]Lingjun Kong, [2]Xiuju Huang, [3]Mengmeng Cao

[1.]School of Electronics & Information, Hangzhou Dianzi University, Hangzhou, China, 310018

[2.]UTSTARCOM Co. Ltd. Hangzhou, China, 310052

[3.]North China Electric Power University, Department of electronics and Communication Engineering, Baoding, China, 071003

Email:[1] islotus@163.com, [2] xjhuang@utstar.com, [3] caomengmeng520@126.com

*Abstract.* By using the dynamic reconfigurable transceiver in high speed interface design, designer can solve critical technology problems such as ensuring signal integrity conveniently, with lower error binary rate. In this paper, we designed a high speed XAUI (10Gbps Ethernet Attachment Unit Interface) to transparently extend the physical reach of the XGMII. The following points are focused: (1) IP (Intellectual Property) core usage. Altera Co. offers two transceiver IP cores in Quartus II MegaWizard Plug-In Manager for XAUI design which is featured of dynamic reconfiguration performance, that is, ALTGX_RECONFIG instance and ALTGX instance, we can get various groups by changing settings of the devices without power off. These two blocks can accomplish function of PCS (Physical Coding Sub-layer) and PMA (Physical Medium Attachment), however, with higher efficiency and reliability. (2) 1+1 protection. In our design, two ALTGX IP cores are used to work in parallel, which named XAUI0 and XAUI1. The former works as the main channel while the latter redundant channel. When XAUI0 is out of service for some reasons, XAUI1 will start to work to keep the business. (3) RTL (Register Transfer Level) coding with Verilog HDL and simulation. Create the ALTGX_RECONFIG instance and ALTGX instance, enable dynamic reconfiguration in the ALTGXB Megafunction, then connect the ALTGX_RECONFIG with the ALTGX instances. After RTL coding, the design was simulated on VCS simulator. The validated result indicates that the packets are transferred efficiently. FPGA makes high-speed optical communication system design simplified.

**Keywords:** *High speed XAUI, Dynamic reconfigurable transceiver, EBR, FPGA*

* Corresponding Author:

Haipeng Zhang

School of Electronics & Information, Hangzhou Dianzi University,

Hangzhou, China, 310018

Email: islotus@163.com

## 1. Introduction

With the development of Internet, the need for higher bandwidth and data rate is driven largely by the demand to upgrade existing communication systems and the emergence of new applications. Communication equipment on market has the trend of lower power and lower cost, higher speed transportation with a more compact size.

The next generation of IC product uses the 45-nm or 40-nm process [1], even the 28-nm process to integrate more functions, so that higher operating performance and higher logic density can be achieved. However, the key to meet the increasing demand for bandwidth is that we use more and faster high-speed serial transceivers. Data links with transceivers support higher data throughput, more power efficient and higher system integration. By using transceiver technology, designers can solve critical technology issues that exist in high speed data link designs, such as Signal Integrity, Power Consumption and Board Complexity.

Altera Corporation announced the availability of a 10G Ethernet (10GbE) reference design targeting designers using the XAUI communications protocol [2], in the aims at addressing the demands of broadband networking and telecommunication applications. Altera offers a broad portfolio of devices with embedded transceivers that support dynamic reconfiguration, both FPGAs and ASICs (Application Specific Integrated Circuit). These are the Arria II GX FPGA family, the Stratix IV GX and GT FPGA families and the HardCopy IV GX ASIC family. They provide a dedicated mode for implementing the XAUI interface and allow the integration of multiple PHYs (Physical) and MACs (Media Access Control) into a single FPGA. The solution enables simple, fast protocol implementation, which reduces design risk, shortens development times, and allows you to concentrate on the core functions of the system design.

## 2. Dynamic reconfigurable architecture

Dynamically reconfiguration means reconfigurable device of the chip can be reconfigured repeatable, and performs different functions at different times [3].So we can get various groups by changing settings of the devices without power off. Compared with static reconfiguration, dynamic reconfiguration can use the reconfigurable device more thoroughly.

### 2.1. Dynamic reconfigurable transceiver introduction

Figure 1 shows the channel architecture of Arria II GX transceiver. Each transceiver channel consists of a transmitter channel and a receiver channel. Each transmitter or receiver channel is composed of one channel PCS block and one channel PMA block [4][5]. Some modules in PCS block are optional, which can be bypassed.

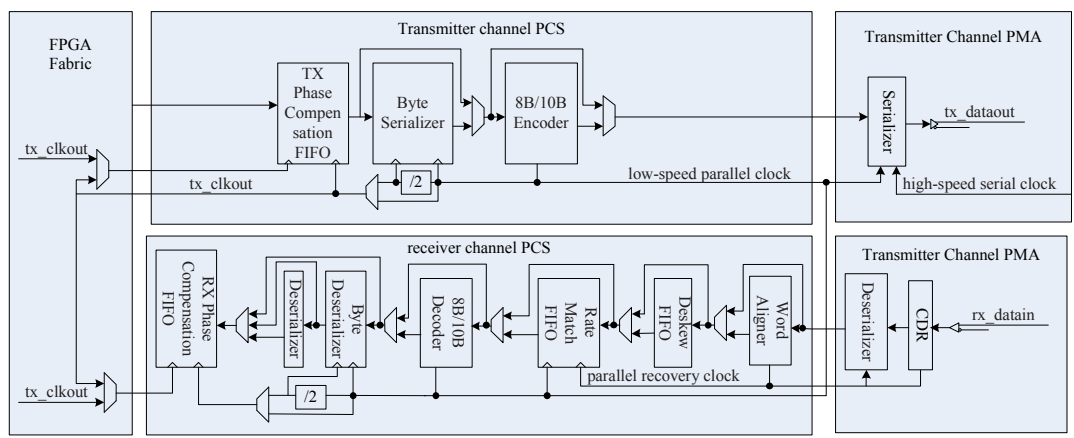

**Figure 1.** Transceiver channel architecture for Arria II GX

Transmitter PCS consists of the TX phase compensation FIFO, byte serializer, and 8B/10B

encoders. While the receiver PCS consists of the word aligner, deskew FIFO, rate-match FIFO, 8B/10B decoder, byte paralyzer, byte ordering, and RX phase compensation FIFO.

Transmitter PMA consists of the serializer and the transmitter output buffer. Receiver PMA has the receiver input buffer, CDR, and paralyzer. The PMA module supports analog interface from the transceiver to medium outside.

## 2.2 Dynamic reconfiguration mode

There are four dynamic reconfiguration modes in Arria II GX devices: Offset Cancellation for receiver channels, Analog Control, Data rate division in transmitter (TX), Channel and TX PLL select/reconfig. Selection of these modes allows you to dynamically reconfigure various channel and clock multiplier units (CMU) settings without powering the device off [6].

Different decisions will be made, which depends on various reconfiguration requirements and application environments when we are choosing reconfiguration mode in the design. **Table 1** lists the reasons to reconfigure the transceivers and the reconfiguration modes for various reconfiguration requirements.

Table 1. Reconfiguration modes for various reconfiguration requirements

| Reason for Reconfiguration | Reconfiguration Mode to Use |
| --- | --- |
| Counter offset variations due to process, voltage, and temperature (PTV) for analog circuits. Mandatory feature if you use receivers. | Offset Cancellation |
| Fine-tune signal integrity by adjusting the transmitter and receiver buffer settings while bringing up a link | Analog Controls |
| Increase or decrease the data rate (/1, /2, /4) for auto negotiation | Data rate division in TX or Channel and TX PLL select/reconfig |
| Support multiple protocols with the same transceivers to add design flexibility | Analog Controls Channel and TX PLL select/reconfig |
| Reset the CMU PLLs through Channel and TX PLL select/reconfig | Channel and TX PLL select/reconfig |

## 2.3 Dynamic reconfiguration architecture

IP cores are used in this design and a technology of high-performance IP core multiplex is chosen. These IP cores must be verified to make sure that they work well. Therefore, design efficiency can be improved if higher reliability and lower design risk is achieved.

Arria II GX device offers two instances to support dynamic reconfiguration: ALTGX_ RECONFIG and ALTGX [7]. **Figure 2** illustrates the connection relationships between them.

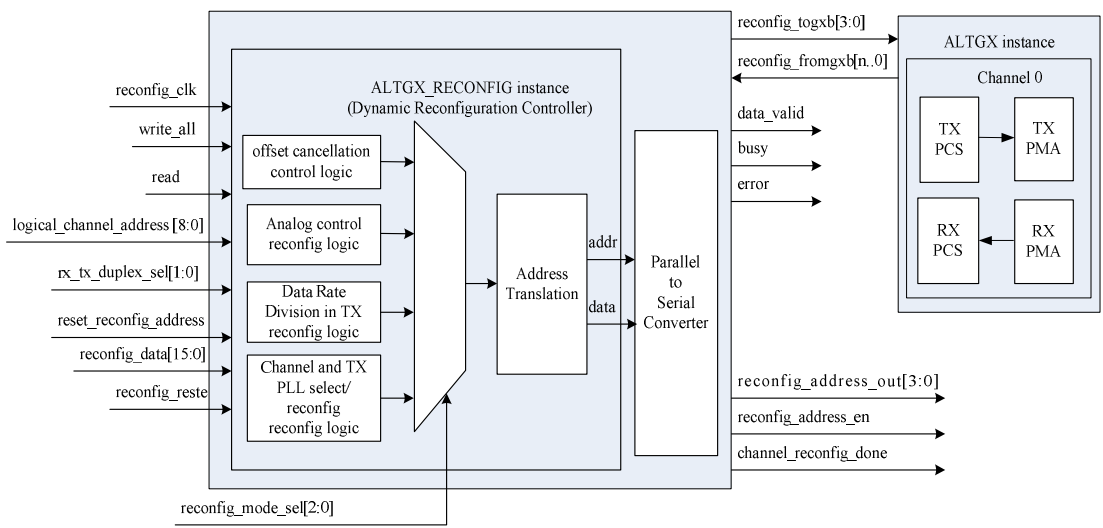

Figure 2. Connection relationships between ALTGX_ RECONFIG and ALTGX

(1) ALTGX_ RECONFIG instance
The ALTGX_RECONFIG instance generated by the ALTGX_RECONFIG MegaWizard Plug-In Manager represents the dynamic reconfiguration controller. It provides simple way to change transceiver PMA settings dynamically.
(2) ALTGX instance
The ALTGX instance generated by the ALTGX MegaWizard Plug-In Manager represents the transceiver. This term is used as the functional module.

## 3 High speed XAUI design

As the demand for bandwidth and transfer rate increases, high speed transceiver plays a more important role in high speed interface design. The design based on dynamic reconfigurable transceiver of FPGA can fit the bill to some extent [8]. Altera and Xilinx have already launched their own FPGA devices which meet the requirements above. With these devices, design of high speed interface becomes more simple, nevertheless, more realizable. And all this will make further promotion in the development of optical fiber communication and wireless devices.

In this part, we designed a high speed XAUI to transparently extend the physical reach of the XGMII, that is, the data output should meet XGMII style.

### 3.1 XAUI brief introduction

XAUI, the 10 Gigabit Attachment Unit Interface, is a technical innovation that dramatically improves and simplifies the routing of electrical interconnections. Developed by the IEEE 802.3ae 10 Gigabit Ethernet Task Force, XAUI delivers 10Gbps of data throughput using four differential signal pairs in each direction [9].

Generally, XAUI is designed as XGMII (10 Gigabit Medium Independent Interface) extended sublayer. The XGMII provides full duplex operation at a rate of 10GbPs between the MAC and PHY. Each direction is independent and contains a 32-bit data path, as well as clock and control signals. In total the interface is 74 bits wide [10]. **Figure 3** shows the data stream between XGMII and XAUI:

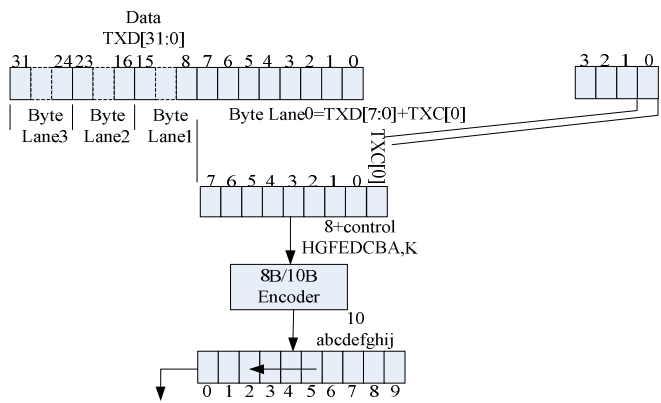

**Figure 3.** Data stream between XGMII and XAUI

The XGMII interface consists of four 8-bit lanes. At the transmit side of the XAUI interface, the data and control characters are converted within the XGMII extender sublayer (XGXS) into an 8B/10B encoded data stream. Each data stream is then transmitted across a single differential pair running at 3.125Gbps [11]. At the XAUI receiver, the incoming data is decoded and mapped back to the 32-bit XGMII format. This provides a transparent extension of the physical reach of the XGMII and also reduces the interface pin count.

### 3.2 High speed XAUI design

The dynamic reconfigurable transceiver embedded in Arria II GX FPGA device is based on 40-nm process. It can provide various kinds of protocols, such as PCI (Peripheral Component Interconnect) Express, serial digital interface, XAUI and so on.

When do research on application of dynamic reconfigurable transceiver in high speed XAUI, we choose the Analog Controls mode. Adjust the parameters of some options, for example, equalizer DC gain, transmit output differential voltage($V_{OD}$), pre-emphasis and setting of equalization, then we will gain the best eye diagram which means the best signal integrity. Something should be noticed that we use receiver channels in the design, so it is necessary to choose Offset Cancellation mode acquiescently. In this section, we will introduce a design proposal of high speed XAUI with dynamic reconfigurable transceiver, and Verilog HDL coding style is used [12]. **Figure 4** illustrates the system top view of the design.

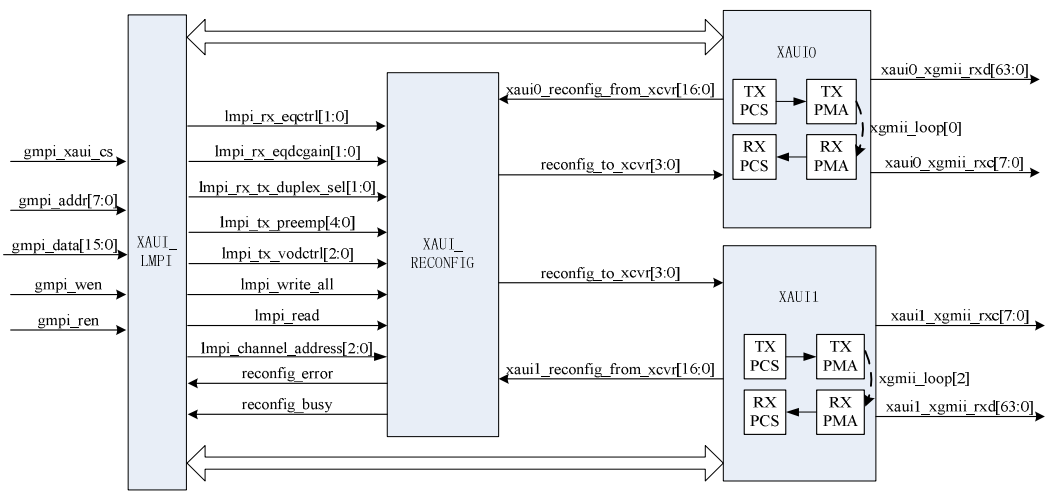

**Figure 4.** System level view of High speed XAUI interface

XAUI_LMPI module functions as the second-level microprocessor interface logic, so the related state and control registers of XAUI can be read and written. Also, we define some registers in this module.

XAUI_RECONFIG module functions as the transceiver controller instance, which is responsible for controlling the high speed XAUI channels. Options in Analog Controls mode have default settings, as the following list: tx_vodctrl is configured to 4, tx_preemp to 0, rx_eqctrl to 0, rx_eqdcgain to 0. Signal lmpi_write_all is used to start the write transportation between XAUI _RECONFIG instance and the XAUI instance, while signal lmpi_read starts the read transportation which is used in PMA control mode only.

XAUI0 and XAUI1 module act as the transceiver instance, implement basic function of high speed XAUI interface. When generate XAUI instance in Quartus II MegaWizard, we set the option "which protocol will you be using" in Parameter Settings to XAUI. Because each channel in the XAUI supports transfer rate up to 3.2Gbps, it is necessary for us to bind 4 such channels together to achieve 10Gbps transfer rate, which means that the option "what is the number of channels" should be set to 4.

### 3.3 Signal description

In the work of RTL coding, we should connect the ALTGX_RECONFIG instance with the ALTGX instances. The most important work is that signals must be connected to the related ports. **Table 2** lists the input and output ports of the XAUI design, **Table 3** lists the internal and external signals of XAUI_RECONFIG.

Table 2. Input and Output Ports of XAUI

| Signal Name | I/O Type | Description |
| --- | --- | --- |
| phy_mgmt_clk | Input | Clock signal |
| rst | Input | Global asynchronous reset signal |
| gmpi_xaui_cs | Input | Chip select signal |
| gmpi_addr[7:0] | Input | Address signal |
| gmpi_data[15:0] | Input | Input data signal |
| gmpi_wen | Input | Write enable signal |
| gmpi_ren | Input | Read enable signal |
| xaui_sel | Input | XAUI bus select control signal:<br>0:XAUI0<br>1:XAUI1 |
| xaui0_xgmii_rxd[63:0] | Output | Internal XGMII received data from XAUI0, signal xaui_sel is set to 0 |
| xaui0_xgmii_rxc[7:0] | Output | Internal XGMII receive control from XAUI0 |
| xaui1_xgmii_rxd[63:0] | Output | Internal XGMII received data from XAUI1 |
| xaui1_xgmii_rxc[7:0] | Output | Internal XGMII receive control from XAUI1 |

Table 3. Internal and External Signals of Dynamic Reconfiguration Controller

| Signal Name | I/O Type | Description |
| --- | --- | --- |
| lmpi_rx_eqdcgain | Input | Equalizer DC gain write control.<br>Legal settings allowed for this signal:<br>2'b00 => 0dB<br>2'b01 => 3dB<br>2'b10 => 6dB<br>All other values => N/A |
| lmpi_rx_eqctrl | Input | Write control to write an equalization control value for the receive side of the PMA.<br>rx_eqctrl[1:0]   Corresponding ALTGX settings<br>2'b00                           0(Low)<br>2'b01                           1<br>2'b10                           2<br>2'b11                           3(High) |
| lmpi_tx_preemp | Input | Pre-emphasis write control for the first post-tap for the transmit buffer.<br>lmpi_tx_preemp[4:0]   Corresponding ALTGX settings<br>5'b00000                           0<br>5'b00001                           1<br>5'b00101                           2<br>5'b01001                           3<br>5'b01101                           4<br>5'b10001                           5<br>5'b10101                           6<br>All other values               N/A |
| lmpi_tx_vodctrl | Input | Transmit buffer $V_{OD}$ control signal.<br>tx_vodctrl           Corresponding ALTGX settings<br>3'b000                           0<br>3'b001                           1<br>3'b010                           2<br>3'b011                           N/A<br>3'b100                           4<br>3'b101                           5<br>3'b110                           6<br>3'b111                           7 |
| lmpi_rx_tx_duplex_sel | Input | Select reconfigured channel.<br>2'b00     The transmitter and receiver portion<br>2'b01     The receiver portion<br>2'b10     The transmitter portion |
| lmpi_write_all | Input | Initiate a write transaction which is asserted for one clock. |
| lmpi_read | Input | Read data signal, asserted for one clock.<br>Only to PMA controls reconfiguration mode. |
| lmpi_channel_address | Input | The width of this signal depends on the value you set in the "What is the number of channels controlled by the reconfigure controller" |
| reconfig_to_xcvr | Output | An output port of ALTGX_RECONFIG instance and an input port of the ALTGX instance. |
| reconfig_error | Output | Indicate that an unsupported operation is attempted. |
| reconfig_busy | Output | Indicate that the busy status of the dynamic reconfiguration controller during offset cancellation mode. |
| xaui0_reconfig_from_xcvr | Input | An output port in the XAUI0. |
| xaui1_reconfig_from_xcvr | Input | An output port in the XAUI1. |

### 3.4 Result analysis

After the design of high speed XAUI based on dynamic transceiver, we did functional simulation by using VCS software from Synopsys. Then further work was done, such as synthesis, placement and routing in Quartus II software [13]. In the test bench file, we use two functions to record the packets information, such as simulation time, packet length, lost packet count etc.
```
    initial begin
        $set_bus(3,2);
        $eth_init(0,4,eth_pkt_num,{eth_ifg,eth_bandwidth},3);
        $set_print(3,1,3,"eth_gen0.dat","eth_chk0.dat");
    end
```
Check the generated file eth_gen0.dat and eth_chk0.dat. File eth_gen0.dat records packets information of packets generated as the upstream input. File eth_chk0.dat details output packets of downstream, and signal xgmii_loop should be set. **Table 4** shows the packets generated from 14c76 ns to 15cca ns, also the packet number and the packet length are printed. **Table 5** lists the received downstream packets at the output port.

Table 4. Upstream Packets Information

| TimeL | ethFCnt | length |
|-------|---------|--------|
| 14c76 | 86 | 588 |
| 150e3 | 87 | 577 |
| 1513d | 88 | 5a |
| 1525d | 89 | 151 |
| 15363 | 8a | 12f |
| 15530 | 8b | 230 |
| 158aa | 8c | 440 |
| 1593d | 8d | a5 |
| 159aa | 8e | 72 |
| 15bf0 | 8f | 2c1 |
| 15cca | 90 | fb |
| 160ea | 91 | 511 |
| 16563 | 92 | 582 |
| 1661d | 93 | d2 |

Table 5. Downstream Packet Information

| TimeL | packetNum | length |
|-------|-----------|--------|
| 1666a | 86 | 588 |
| 16ad6 | 87 | 577 |
| 16b2a | 88 | 5a |
| 16c43 | 89 | 151 |
| 16d4a | 8a | 12f |
| 16f16 | 8b | 230 |
| 1728a | 8c | 440 |
| 17316 | 8d | a5 |
| 17383 | 8e | 72 |
| 175c3 | 8f | 2c1 |
| 1769d | 90 | fb |
| 17ab6 | 91 | 511 |
| 17f30 | 92 | 582 |
| 17fe3 | 93 | d2 |

These two tables illustrate that the input packets are received at the output port after a reasonable time's transmission, with the same length. In addition, eth_chk0.dat prints the lost packets count and the transmission status, so it is easy to debug.

In order to confirm the function of XAUI, we captured a screenshot on VCS simulator. **Figure 5** shows the verification result of the output packets at time 15300ns, xgmii_txd[63:0] is the input signal, while xgmii_rxd[63:0] is the output signal of XAUI, and xgmii_txc[7:0] is the control byte.

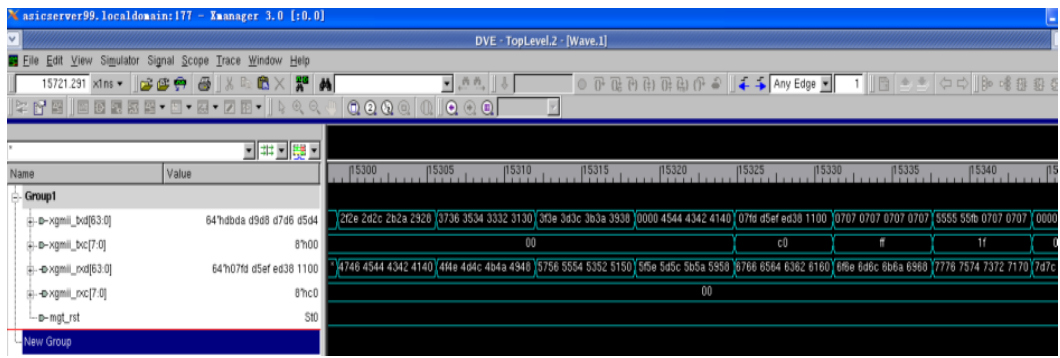

**Figure 5.** Transmitted and received packets at time 15300ns

When TXC is 0, xgmii_txd[63:0] present normal data transmission. When TXC is 1, if FB byte is detected in xgmii_txd[63:0], a packet start to transmit. If FD is detected, a packet transmission is at the end.

The designed XAUI module is used on TN725 devices which developed by UTSTARCOM Co. Ltd. It is used to complete the conversion between XAUI on back panel and the internal XGMII. Though the design meets requirement quite well, it has its own limitations. First, IP cores from Altera are needed, we have to get permission from Altera. Second, timing is very important in FPGA design, meeting the timing requirements was really a hard work.

## 4. Conclusions

In the design of the proposed high speed XAUI, dynamic reconfigurable transceiver allows us to set related parameters from outside registers, which makes the best eye diagram practicable. By virtue of IP core usage, RTL coding in FPGA and SOC design becomes easy and reliable in many research areas [14]. However, there are several parameters to set, such as lmpi_tx_preemp[4:0] and lmpi_tx_vodctrl[2:0], so it's really an arduous task which increased difficulties. In the design, two authenticated XAUI cores are adopted which work as 1+1 protection. As a result, CPLD can generate selective signal automatically according to the position of the two PPC boards. If one of them failed to work, another will work well to keep data transmission. Though great reliability can be achieved in this method, disadvantages followed. It will reduce the channel utilization. After the process of synthesis with Quartus II 11.0, we find that 1924 LUTs (Look Up Table) of the FPGA are consumed. As a result, the design method based on IP cores is benefit to cut down design cycle, shorten press time to market, improve design efficiency and increase design reliability.

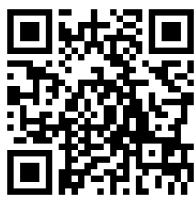

Free download this article and more information